\documentclass[a4paper,fleqn]{cas-dc}

\usepackage[numbers]{natbib}
\usepackage{float}
\usepackage{xfrac}
\usepackage{algorithm}
\usepackage[noend]{algpseudocode}
\usepackage{graphicx}
\usepackage{blindtext}
\usepackage{afterpage}
\usepackage[export]{adjustbox}
\usepackage{multicol}
\usepackage[utf8]{inputenc}
\usepackage{amssymb,amsfonts}
\everymath{\displaystyle}
\usepackage[many]{tcolorbox}
\usepackage[figuresright]{rotating}
\usepackage{array}
\usepackage{kantlipsum}
\usepackage{gensymb}
\usepackage{xpatch}
\usepackage{tikz}
\usepackage{lipsum}
\usepackage{multirow} 
\usepackage{lineno}
\usepackage{amsmath}
\usepackage{textcomp}
\usepackage{amsbsy,enumerate}
\usepackage{ccaption}
\usepackage{chemformula}
\usepackage{balance}
\usepackage{ragged2e}
\usepackage{setspace}
\usepackage{subcaption}
\usepackage{etoolbox}
\usepackage{siunitx}
\usepackage{mathtools}
\usepackage{xcolor}
\usepackage{fullpage,anysize,esint,relsize, commath}
\usepackage{fancyhdr} 
\usepackage{hyperref}
\usepackage{enumitem}
\usepackage{pifont}
\usepackage{tabularray}
\usepackage{listings}
\usepackage{cprotect}
\usepackage{multiaudience}
\usepackage{abbrevs}
\usepackage{isomath}
\usepackage{graphics}
\usepackage{stfloats}
\usepackage{geometry}
\usepackage[cjk]{kotex}
\usepackage{longtable}
\usepackage{appendix}

\DeclareMathAlphabet{\mathpzc}{OT1}{pzc}{m}{it}
\DeclareMathAlphabet{\mathbbm}{U}{bbm}{m}{n}

\def\tsc#1{\csdef{#1}{\textsc{\lowercase{#1}}\xspace}}
\tsc{WGM}
\tsc{QE}
\tsc{EP}
\tsc{PMS}
\tsc{BEC}
\tsc{DE}
\begin{document}

\algnewcommand\algorithmicswitch{\textbf{switch}}
\algnewcommand\algorithmiccase{\textbf{case}}
\algnewcommand\algorithmicassert{\texttt{assert}}
\algnewcommand\Assert[1]{\State \algorithmicassert(#1)}%
% New "environments"
\algdef{SE}[SWITCH]{Switch}{EndSwitch}[1]{\algorithmicswitch\ #1\ \algorithmicdo}{\algorithmicend\ \algorithmicswitch}%
\algdef{SE}[CASE]{Case}{EndCase}[1]{\algorithmiccase\ #1}{\algorithmicend\ \algorithmiccase}%
\algtext*{EndSwitch}%
\algtext*{EndCase}%

\let\WriteBookmarks\relax
\def\floatpagepagefraction{1}
\def\textpagefraction{.001}
\title [mode = title]{CNN-LSTM and Transfer Learning Models for Malware Classification based on Opcodes and API Calls}
\author[1]{Ahmed Bensaoud}
\cormark[1]
\ead{abensaou@uccs.edu}
\author[1]{Jugal Kalita}
\ead{jkalita@uccs.edu}
\address[1]{Deptarment of Computer Science, University of Colorado Colorado Springs}
\cortext[cor1]{Corresponding author:
Tel.: +1-970-581-6683;}
 
\begin{abstract}
In this paper, we propose a novel model for a malware classification system based on Application Programming Interface (API) calls and opcodes, to improve classification accuracy. This system uses a novel design of combined Convolutional Neural Network and Long Short-Term Memory. We extract opcode sequences and API Calls from Windows malware samples for classification. We transform these features into N-grams (N = 2, 3, and 10)-gram sequences. Our experiments on a dataset of  9,749,57 samples produce high accuracy of \textbf{99.91\%} using the 8-gram sequences. Our method significantly improves the malware classification performance when using a wide range of recent deep learning architectures, leading to state-of-the-art performance. In particular, we experiment with ConvNeXt-T, ConvNeXt-S, RegNetY-4GF, RegNetY-8GF, RegNetY-12GF, EfficientNetV2, Sequencer2D-L, Swin-T, ViT-G/14, ViT-Ti, ViT-S, VIT-B, VIT-L, and MaxViT-B. Among these architectures, Swin-T and Sequencer2D-L architectures achieved high accuracies of 99.82\% and 99.70\%, respectively, comparable to our CNN-LSTM architecture although not surpassing it. 
\end{abstract}
\maketitle
\section{Introduction}
Malware is malicious code that enters a computer or an internet-connected device to steal sensitive information from government, commercial or private organizations. Internet-connected devices infected with malware can also destroy and/or gain access to confidential information, randomly reboot, track user activity, make devices run slower, start unknown processes, or send emails without user action. Classifying malware is complicated because most malware developers adopt strategies to avoid anti-virus systems. Therefore, it is important to fight common and modern evasion techniques to improve the analysis results. Reverse engineering helps understand how a malware works by monitoring runtime execution using dynamic analysis tools, giving insightful information. 

Reverse engineering is the process of analyzing software in order to understand its design, functionality, and behavior. This technique is often used in malware analysis to identify and understand the nature of malicious code. Some examples of how reverse engineering is used in malware analysis are given below.
\begin{itemize}
    \item \textbf{Disassembling code}: One of the primary techniques used in reverse engineering is disassembling the code of a malware sample. This process involves converting the binary code of the malware into human-readable assembly language, which can then be analyzed to understand the behavior of the malware.
\item \textbf{Malware Analysis}: Another important technique in reverse engineering is malware analysis, which involves running the malware in a controlled environment and analyzing its behavior as it executes. This can help identify the specific functions and routines used by the malware, as well as any malicious behavior it exhibits.
\item \textbf{Developing countermeasures}: Reverse engineering can also be used to develop countermeasures to malware. By analyzing the code of a malware sample, researchers can identify its specific characteristics and develop tools and techniques to detect and remove it from infected systems.
\end{itemize}
Overall, reverse engineering is an important technique in  malware analysis, helping researchers understand the behavior of malicious code and develop effective countermeasures to protect against it. In January 2023, researchers discovered that malware authors had begun using VSCode malware extensions as a new attack vector to launch malicious attacks. These extensions can take many forms, such as adware, spyware, or even ransomware. Reverse engineering techniques can be used to analyze the code of these extensions in order to understand their behavior and develop effective countermeasures\footnote{https://www.esecurityplanet.com/threats/vscode-security}. Malware authors have created a new Rootkit malware that harbors JavaScript code that, when launched, paves the way for additional payloads such as SNOWCONE Cobalt Strike, FONELAUNCH, and Beacon\footnote{https://thehackernews.com/2023/01/gootkit-malware-continues-to-evolve.html}.  

Additionally, Application Programming Interfaces (API) allow different software applications or systems to communicate and interact with each other. Malware can exploit API vulnerabilities to execute unauthorized actions, manipulate data, or gain access to sensitive information. An opcode is a part of the machine code, which is a low-level representation of instructions that a computer processor can execute. Malware often uses specific sequences of opcodes to carry out malicious activities, such as gaining unauthorized access, stealing sensitive information, or spreading further within a system. Advanced machine learning algorithms can be used to analyze massive amounts of API call data and establish baseline behaviors for normal API interactions. Machine learning models can then be trained to detect deviations from these baselines, helping identify and prevent malware attacks. They can be trained to recognize opcode patterns that are indicative of malware. By analyzing large datasets of opcode sequences, these algorithms can learn to detect new and emerging threats that may not have known signatures.

Over the last few years, researchers have developed various approaches to classify malware using text (code) or images. Recently, methods from computer vision, machine learning , deep learning , and transfer learning have been used to detect malware automatically \cite{bensaoud2020classifying,yoo2021ai,aslan2021new,he2022deep}. In particular, Deep learning is used as a feature extractor that enhances classification accuracy \cite{dib2021multi}. 

Transfer learning can be effective for malware image classification tasks. Transfer learning involves taking a deep learning model that has been pre-trained on a large dataset of non-malware images (malware files in binary 2-D format arranged in a matrix like an image) and fine-tuning it on a smaller dataset of malware images. By doing so, the model can learn to classify malware images with high accuracy without requiring as much labeled data. One approach to employing transfer learning for malware image classification involves using a pre-trained convolutional neural network (CNN) as a feature extractor. The CNN is trained on a large dataset of non-malware images, and its weights are frozen. The malware images are then passed through the CNN to obtain feature vectors, which are then used to train a classifier. The classifier can be a simple linear classifier or a more complex model, such as a support vector machine (SVM) or a random forest. Another approach is to \textbf{fine-tune} the entire pre-trained CNN on the malware images. This involves unfreezing some or all of the layers of the pre-trained CNN and training them on the malware images while also updating the weights of the classifier. This approach can be more effective than using the pre-trained CNN as a feature extractor since the entire model can be optimized for the malware classification task.

Additionally, exploiting Application Programming Interface (API) vulnerabilities and analyzing opcodes (machine code instructions) offer insights into malware behavior. Advanced machine learning algorithms, particularly those employing deep learning architectures, have proven effective in analyzing massive datasets of API calls and opcode sequences to detect and prevent malware attacks. %The literature has witnessed a shift toward using methods from computer vision, machine learning, transfer learning, and deep learning for automated malware detection.

Our paper introduces a novel approach to classify malware, focusing on opcode sequences and API calls extracted from diverse malware samples. The main contributions of this paper include:

\begin{itemize}
\item
\textbf{Innovative Classification Model:} We propose a novel model for classifying malware families, incorporating Convolutional Neural Network (CNN), Long Short-Term Memory (LSTM), and techniques inspired by Natural Language Processing (NLP). This innovative model aims to enhance classification accuracy. In addition, our approach to representing API calls and opcode sequences is novel, further improving the results.

\item \textbf{Empirical Evaluation:} We conduct extensive experiments to assess the performance of various fine-tuned recent pre-trained deep learning models, including ConvNeXt-T \cite{liu2022convnet}, ConvNeXt-S \cite{liu2022convnet}, RegNetY-4GF\cite{radosavovic2020designing}, RegNetY-8GF\cite{radosavovic2020designing}, RegNetY-12GF\cite{radosavovic2020designing}, EfficientNetV2 \cite{leng2022polyloss}, Sequencer2D-L \cite{tatsunami2022sequencer}, ViT-G/14 \cite{wortsman2022model}, ViT-Ti \cite{strudel2021segmenter}, ViT-S \cite{strudel2021segmenter}, VIT-B \cite{dehghani2021efficiency}, VIT-L \cite{dehghani2021efficiency}, MaxViT-B \cite{tu2022maxvit}, and Swin-T \cite{liu2021swin}. Our evaluations provide empirical evidence of the effectiveness of our proposed CNN-LSTM approach with innovative representation for API calls and opcode sequences, showcasing advancements in malware classification.
\end{itemize}
Both contributions collectively aim to advance the field of malware classification by introducing innovative approach, providing empirical evidence of effectiveness and highlighting how our approach compares with a large number of very recent methods. The remainder of this paper is organized as follows: Section 2 discusses related work, Section 3 details feature extraction techniques, Section 4 explains the methodologies, Section 5 presents the dataset, Section 6 details the experiments and results, Section 7 provides a short analysis, and the paper concludes with Section 8, which includes the references. All notations used in the paper are listed in Table 13.

\section{Related work}
In the realm of malware classification, the combination of Convolutional Neural Networks (CNNs) and Long Short-Term Memory (LSTM) networks has emerged as a powerful strategy. CNNs excel in extracting latent features from non-sequential data, particularly images, while LSTMs are adept at capturing dependencies within sequential data, making them invaluable for classification and prediction tasks. The proposition of a CNN-LSTM model for malware classification stems from the synergistic advantages these models offer. CNNs contribute by filtering noise and extracting crucial features from input data, while LSTMs efficiently capture intricate sequence patterns. This strategic amalgamation exploits the strengths of both deep learning approaches, resulting in a notable enhancement in malware classification performance.

\subsection{CNN-LSTM Models for Malware Classification}
\citet{Zhang9148385} introduced a CNN-LSTM framework for malware classification by extracting features from n-grams of API calls. \citet{peng2019malicious} proposed an Attention-Based CNN-LSTM model for detecting malicious URLs, achieving 98.18\% accuracy. \citet{sun2020dl} developed a CNN-LSTM model for intrusion detection using network traffic data, achieving 98.67\% accuracy on the CICIDS2017 dataset. \citet{kuang2019deepwaf} designed DeepWAF to detect web attacks in HTTP requests using CNN-LSTM. \citet{praanna2020cnn} presented a CNN-LSTM model for intrusion detection, leveraging spatial and temporal features for improved performance.

\subsection{Transfer Learning for Malware Classification}
\citet{garcia2023effectiveness} proposed a method to evaluate the effectiveness of transfer learning techniques in malware detection, contributing valuable insights. \citet{CHAGANTI2022103306} introduced an EfficientNetB1-based malware classification approach, achieving a remarkable 99\% accuracy on the Microsoft Malware Classification Challenge (MMCC) dataset with fewer parameters. \citet{khan2019analysis} conducted a comprehensive analysis of older pre-trained models (Inception-V4, ResNet18, ResNet34, ResNet50, ResNet101, ResNet152) for malware classification on the MMCC dataset, and highlighted the superior performance of ResNet152 with a testing accuracy of 88.36\%. \citet{ullah2022explainable} employed the Bidirectional Encoder Representations from Transformers (BERT) model \cite{devlin2018bert} for feature extraction, introducing a unique malware-to-image conversion algorithm. They utilized the FAST (Features from Accelerated Segment Test) \cite{viswanathan2009features} extractor and BRIEF (Binary Robust Independent Elementary Features) descriptor \cite{calonder2010brief} to efficiently extract and emphasize significant features. The trained and texture features were then combined and balanced using the Synthetic Minority Over-Sampling (SMOTE) \cite{chawla2002smote} method and a CNN network was employed to extract deep features. The study employed a balanced ensemble model, incorporating CNN networks for deep feature extraction, leading to effective malware classification and detection.\\

This paper not only underscores the significance of CNN-LSTM models in malware classification, but also highlights recent advancements in the integration of transfer learning techniques, demonstrating how to further push the boundaries of efficacy in malware detection.

\section{Feature Extraction Techniques}
In order to use the transfer learning models as well as a custom CNN-LSTM model we propose, we need to extract features of malware. Two feature extraction techniques, Bag of Words (BoW), and TF-IDF are used. One-hot-encoding is used to represent categorical features. 

\subsection{Term Frequency-Inverse Document Frequency (TF-IDF)}
Term Frequency-Inverse Document Frequency (TF-IDF) is a statistical measure used to evaluate the importance of a word in a document. It is commonly used in text mining and information retrieval.\\

The formula for TF-IDF is:

\begin{equation}\label{eq:equation1}
TF\mbox{-}IDF(t,d) = tf(t,d) \cdot idf(t).
\end{equation}

\color{blue}{In equation ~\ref{eq:equation1}}\color{black}, $tf(t,d)$ is the frequency of term $t$ in document $d$,
$idf(t)$ is the inverse document frequency of term $t$ across all documents in the corpus. In particular, the term frequency $tf(t,d)$ is the number of times term $t$ appears in document $d$. In our case, a document is the text of the code in an API and opcodes as well. The inverse document frequency $idf(t)$ is calculated as:

\begin{equation}\label{eq:equation2}
idf(t) = log\frac{N}{df(t)}.
\end{equation}

\color{blue}{In equation~\ref{eq:equation2}}\color{black}, $N$ is the total number of documents in the corpus, and $df(t)$ is the number of documents in the corpus that contain term $t$. Logarithmic scaling is used to prevent the bias towards the commonly occurring terms. TF-IDF gives higher weight to terms that appear frequently in a particular document, but are rare across all documents. The total number of documents refers to the total number of malware samples in our dataset.

\subsection{Bag of Words (BoW)}
A bag of words represents the frequency of occurrence of each unique word within a document, without considering semantic or grammatical knowledge. BoW involves representing malware samples as a collection of unique words or features extracted from their code or associated metadata. BoW can also be used for the construction of feature vectors for deep learning algorithms.

\subsection{Concatenation of BoW and TF-IDF}
We simply concatenate TF-IDF features and BoW features for malware as shown in Figure 1. 

\begin{figure*}[ht]
\begin{center}
	\includegraphics[width=15cm, height=14cm,keepaspectratio, frame]{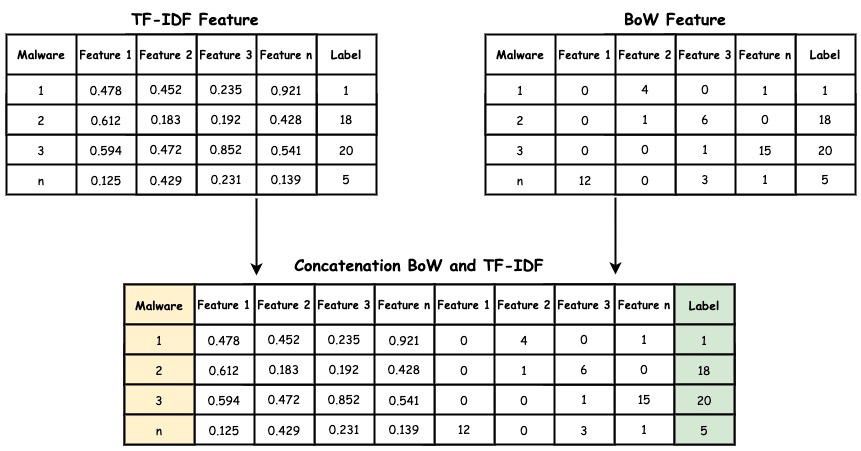}
 \end{center}
	\caption{Obtain TF-IDF and BoW for each malware sample and then concatenate.}
\end{figure*}

\subsection{N-Gram Representation}
The n-gram representation keeps the count of various sequences of n opcodes/API calls. We can represent malware features based on the idea of n-grams. We can set different n-values $( n = 1 , 2 , … , N )$ to run various experiments. Experiments are conducted with $n= 2 .. 10$.

\subsection{One-Hot Encoding} 
One-Hot Encoding is a technique used to transform categorical features into numeric that can be used for machine learning algorithms. This technique counts the unique values and assigns a unique index to each value.

\subsection{API Collection and Opcode}
We extract the API calls and opcodes using two virtual environments, Postman\footnote{\url{https://www.postman.com}} and SoapUI \footnote{\url{https://www.soapui.org}}. The goals for using these tools are to extract Native APIs such as low-level APIs or undocumented APIs, thus going beyond regular APIs and get an extensive number of features. All Native APIs begin with the "Zw" or "Nt" prefixes, e.g., NtRemoveProcessDebug and ZwCreateFile.

\subsection{Transfer learning}
Transfer learning (TL) uses a model that has been pre-trained on one task with a lot of data that can be fine-tuned and used for a different task that does not have a lot of data to train. We evaluate the transfer learning capabilities of various CNN and Transformer architectures and compare the accuracy of classification of the malware dataset. This research uses fourteen recent pre-trained models: ConvNeXt-T \cite{liu2022convnet}, ConvNeXt-S \cite{liu2022convnet}, RegNetY-4GF \cite{radosavovic2020designing}, RegNetY-8GF \cite{radosavovic2020designing}, RegNetY-12GF \cite{radosavovic2020designing}, EfficientNetV2 \cite{leng2022polyloss}, Sequencer2D-L \cite{tatsunami2022sequencer}, ViT-G/14 \cite{wortsman2022model}, ViT-Ti \cite{strudel2021segmenter}, ViT-S \cite{strudel2021segmenter}, VIT-B \cite{dehghani2021efficiency}, VIT-L \cite{dehghani2021efficiency}, MaxViT-B \cite{tu2022maxvit}, and Swin-T \cite{liu2021swin} for malware classification. By transferring the knowledge from these models, even with limited labeled data, one can still achieve good classification performance on malware data. All these models offer a practical and effective solution for classification tasks by leveraging pre-trained models' learned representations, reducing training time and resource requirements, and improving generalization and robustness on limited labeled data. 

\subsubsection{Vision Transformers (ViTs)}
Vision Transformers (ViTs) have achieved state-of-the-art image classification performance using self-attention.
ViTs utilize the Transformer architecture for computer vision tasks. ViTs \cite{dosovitskiy2020image, touvron2021training, wang2021pyramid} show excellent results on the ImageNet classification. The Vision Transformer has developed rapidly in recent years, with a number of variants such as ViT-Ti, ViT-S, VIT-B, VIT-L, Swin-T, MaxViT-B to the recent ViT-G/14. 

\textbf{ViT-Ti} \cite{strudel2021segmenter}, \textbf{ViT-S} \cite{strudel2021segmenter} , \textbf{VIT-B} \cite{dehghani2021efficiency} , and \textbf{VIT-L} \cite{dehghani2021efficiency} based Transformers incorporate self-attention on patch-level information to perform better feature extraction.

The \textbf{Swin-T} \cite{liu2021swin} Transformer is composed of an even number
of Transformer blocks that replace the standard multi-head self-attention module (MSA) with a shifted-window
multi-head self-attention (SW-MSA) and window multi-head self-attention (W-MSA). Swin-T aims to solve problems of computational complexity and lack of information interaction between groups of Transformer layers. Swin-T consists of four stages, where each stage reduces the resolution of the input feature map to expand the approachable field layer by layer. Swin-T can automatically learn from an n-dimensional data matrix and find discriminative and representative features and obtain final classification results.

The Transformer learns global features, but lacks inductive bias and often overfits the training set, unlike CNN, which learns inductive bias but does not learn global features \cite{han2021transformer}. However, \textbf{MaxViT-B} can learn local features at the same time as learning global features \cite{tu2022maxvit}. 

\subsubsection{ConvNeXt-T \& ConvNeXt-S}

ConvNeXt was proposed by Facebook AI Research (FAIR) and UC Berkeley in 2022. ConvNeXt-T \cite{liu2022convnet} and ConvNeXt-S \cite{liu2022convnet} are variations of the ConvNeXt CNN architecture. ConvNeXt-T is a smaller and lighter version, suitable for resource-constrained environments, while ConvNeXt-S provides a balance between model size and performance. Both models leverage grouped convolutions to efficiently capture spatial and channel relationships, making them effective choices for various classification tasks.

\subsubsection{RegNetY-4GF \& RegNetY-8GF \& RegNetY-12GF}
The RegNet \cite{radosavovic2020designing} type models impose a restriction that there is a linear parameterization of block widths, where the block is a modular unit based on the standard residual bottleneck block with group convolution:
\begin{equation}\label{eq:equation3}
u_j=w_0+w_{\alpha}.j.
\end{equation}
\color{blue}In equation~\ref{eq:equation3},\color{black} 
\hspace{2mm} $u_j$ represents the width of the block at index $j$ within the range $j < d$, where $d$ represents the depth of the network. The parameter $w_0 > 0$ signifies a base width for the block, while $w_{\alpha} > 0$ determines the slope of the linear relationship, influencing the width increases with the depth index $j$. RegNetX \cite{radosavovic2020designing} has an additional restriction which sets the bottleneck ratio $b=1$, $12 \le d \le 28$ and $w_m \ge 2$,
where $w_m$ is the width multiplier. RegNetY \cite{radosavovic2020designing} is a fast network that uses residual bottlenecks with a group of simple convolutional models that use the Squeeze-and-Excitation \cite{hu2018squeeze} operation.  Squeeze-and-Excitation improves the strength of a network by explicitly modeling the interdependencies between the channels of its convolutional features. 

\subsubsection{EfficientNetV2}
EfficientNetV2 \cite{leng2022polyloss} is a family of convolutional neural network architectures designed to achieve high performance while being computationally efficient. It is an evolution of the original EfficientNet \cite{tan2019efficientnet} models and introduces several improvements and advancements. EfficientNetV2 incorporates a compound scaling method that uniformly scales up the network's depth, width, and resolution to achieve better accuracy. It also introduces a new model scaling technique called "CoDA" (Compound Domain Adaptation) \cite{chen2012marginalized} that further enhances performance across different domains.

\subsubsection{Sequencer2D-L}
The Sequencer2D-L \cite{tatsunami2022sequencer} model uses LSTMs rather than self-attention layers where the LSTMs are arranged as vertical and horizontal LSTMs to enhance performance. The specific architectural details and optimizations of Sequencer2D-L may vary, but the overarching idea revolves around utilizing LSTMs to effectively transfer learned representations from pre-training tasks to new sequential data in a transfer learning setting.

\section{Methodology}

This study proposes a CNN-LSTM model, in addition to fine-tuning of the state-of-the-art transfer learning models discussed above for modern malware classification. We compare results produced by these classifiers.\\

In our feature engineering strategy, we discovered that the use of 8-gram feature vectors for API calls and opcodes produces an effective integration of advanced techniques. Specifically, we employ natural language processing methods such as bag-of-words (BOW), Term Frequency-Inverse Document Frequency (TF-IDF), and One-hot Encoding as discussed in Subsections 3.1-3.5. These techniques are carefully chosen to capture the diverse nuances of data characteristics, with each method contributing to enhance the representation of the dataset.

The resultant feature vectors, identified as X and Y, serve crucial roles in strengthening our model's discernment capabilities. X, generated through the One-hot Encoding of n-grams, produces a feature vector that clearly outlines the presence of specific sequences of API calls within the dataset. Concurrently, Y takes shape as a feature vector resulting from the intentional concatenation of BOW and TF-IDF representations of grams, giving a richer representation on individual API calls in context.

This purposeful fusion combines the simplicity of BOW with the nuanced representational capability derived from TF-IDF's weighted approach, fostering a comprehensive understanding of the data. The interplay of X and Y within our model not only ensures a holistic grasp of intricate patterns, but also elevates the adaptability and sophistication of our CNN-LSTM architecture and our fine-tuned transfer learning architectures, positioning them as effective solutions for landscape of malware classification. This refined feature engineering methodology, with its incorporation of diverse representations, underscores our model's ability to address the complexity of malware classification, offering a robust and versatile solution.

To convert the concatenated representation to an n-dimensional space, let us denote the Bag-of-Words (BoW) representation as BoW($f$) for a file $f$ containing API calls and opcodes, and the Term Frequency-Inverse Document Frequency (TF-IDF) representation as TFIDF($f$), which also contains API calls and opcodes. The data come from 8-grams, and we convert the concatenated representation to an n-dimensional space by concatenating BoW and TF-IDF representations for each file as follows:\\
\[
\begin{split}
\text{Concatenated}(f)_n = [&\text{BoW}(f)_1, \text{BoW}(f)_2, \ldots, \text{BoW}(f)_C, \\
&\text{TFIDF}(f)_1, \text{TFIDF}(f)_2, \ldots, \text{TFIDF}(f)_C]
\end{split}
\]
\\
$X=\text{Concatenated}(f)_n$\\
where $C$ is the size of the opcode and API call. We assume $n$ is the dimensionality of the concatenated feature vector. This is shown in Figure 1.\\

Additionally, we introduce a new feature \( Y \) where \( Y \) is constructed as a One-Hot Encoding from the 8-grams. This means that for each 8-gram in the dataset, \( Y \) will have a binary entry indicating its presence or absence in a particular malware file.

\subsection{Custom CNN-LSTM Model}

The CNN-LSTM model is designed to effectively classify malware families based on the abstraction and representation of 8-gram sequences involving API calls and opcodes. The model employs a combination of Convolutional Neural Network (CNN) and Long Short-Term Memory (LSTM) networks to automate the generation of sequential feature maps for this specific malware classification task.

The CNN was used to extract complex features from the 8-dimensional matrix, and LSTM was used for classification. The model has three convolutional layers, three pooling layers, three LSTM layers, one fully connected layer. The size of the convolutional layer used for feature extraction is 9 × 9. We use the ELU activation function \cite{clevert2015fast}. Max-pooling kernels of size 3 × 3 are used to reduce the dimensions of the feature maps. Subsequently, a flatten layer is used to transform the output into a one-dimensional vector. At the end of the CNN, the feature map is transferred to the LSTM layer. The model architecture includes an LSTM layer with 512 neurons, a depth of 3, and a dropout rate of 0.3. This LSTM component enhances the model's capacity to discern intricate patterns within the sequential data. The proposed architecture that mixes two deep learning models is shown in Figure 2.

The optimization process for the model involved 200 epochs and a batch size of 64 during training. The Adam optimizer, featuring a learning rate of 0.001, was preferred for its adaptive learning rate calculations, making it a suitable choice for efficient optimization. This combination of architectural intricacies with especially constructed features, underscores the model's robustness and effectiveness in malware classification. The pseudocode for CNN-LSTM-3 is given in Algorithm 1.

In our experimentation, we compared GRU, LSTM, RNN, and CNN models with the proposed hybrid CNN-LSTM-3 architecture for malware classification, utilizing opcodes and API calls as input features. The objective was to thoroughly assess and compare the performance of the proposed model against a number of recent high-performing CNN models.

\begin{figure*}[ht]
\begin{center}
	\includegraphics[width=17cm, height=15cm,keepaspectratio, frame]{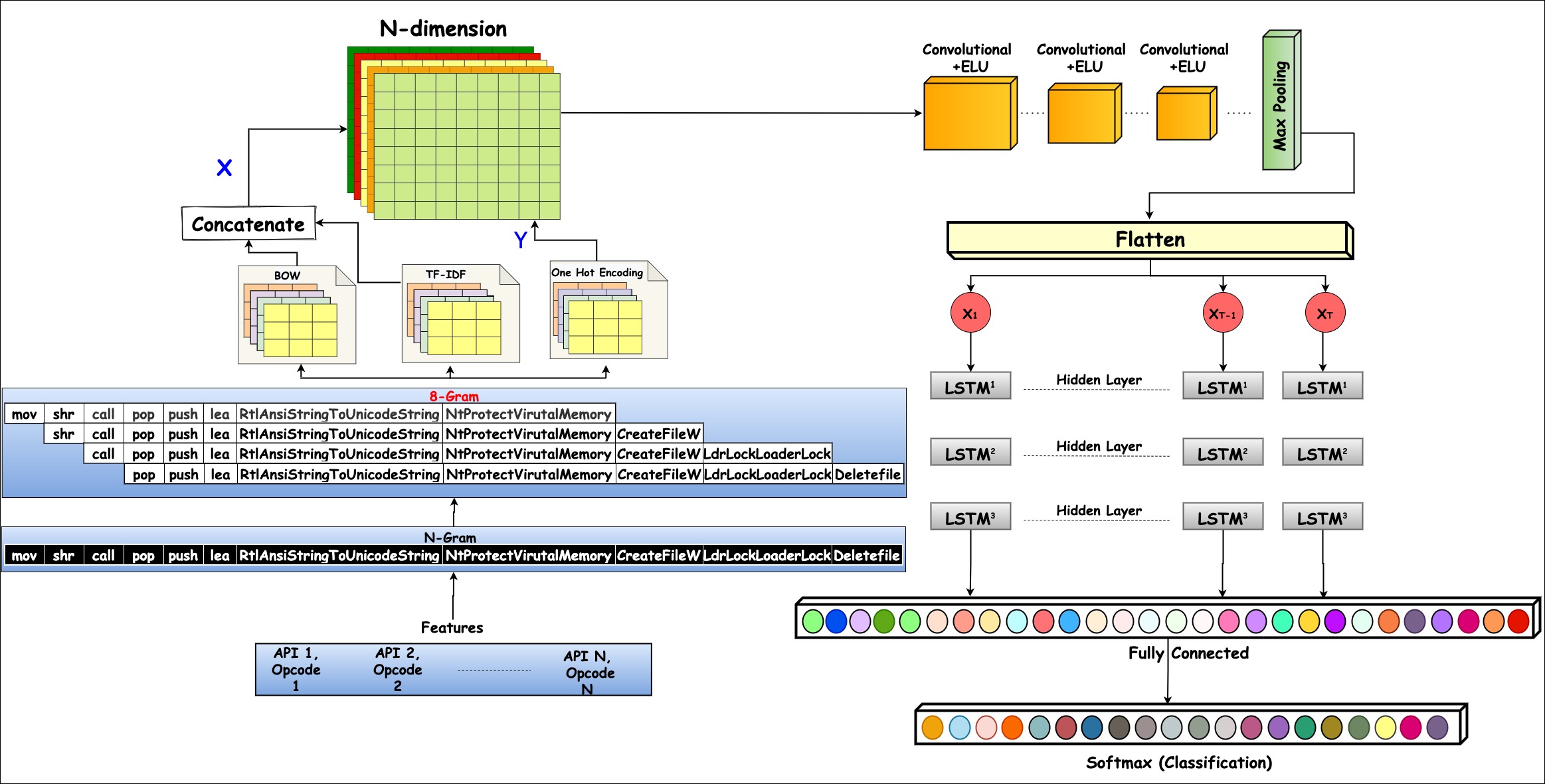}
 \end{center}
	\caption{Proposed model of CNN-LSTM.}
\end{figure*}

\subsection{Fine-tuned State-of-the-art Models}
In second proposed approach, we adopted the same input structure that was used in the previous model and applied it to various transfer learning models, as shown in Figure 3.

Fine-tuning is a crucial component of the transfer learning approach, allowing for reuse of a selection of pre-trained layers to enhance overall performance. In our endeavor to improve the accuracy of pre-trained models, we implemented specific fine-tuning configurations. The models discussed earlier in Section 3.7 have been fine-tuned for the classification of 20 malware families.\\

All models are fine-tuned by adjusting the pre-trained weights to suit the specific task. \\
Let $\theta$ denote the parameters of a pre-trained model, and $\mathcal{L}$ represent the loss function used for training. The standard objective is to minimize this loss as shown \color{blue}{in equation~\ref{eq:equation4}}\color{black}:
% Fine-tuning objective
\begin{equation}\label{eq:equation4}
\text{Fine-tuning objective:} \quad \underset{\theta}{\text{minimize}} \quad \mathcal{L}(\theta).
\end{equation}
During fine-tuning, the model is typically trained on a new dataset related to the malware families. Let $\mathcal{D}_{\text{new}}$ represent this dataset, and $\mathcal{L}_{\text{new}}$ denote the corresponding loss function as illustrated \color{blue}{in equation~\ref{eq:equation5}}\color{black}:
% New loss function for the fine-tuned model
\begin{equation}\label{eq:equation5}
\mathcal{L}_{\text{new}}(\theta) = \frac{1}{|\mathcal{D}_{\text{new}}|} \sum_{(x, y) \in \mathcal{D}_{\text{new}}} \mathcal{L}(\theta, x, y).
\end{equation}
Fine-tuning involves updating the weights ($\theta$) based on the gradients of the loss with respect to the parameters, as outlined \color{blue}{in equation ~\ref{eq:equation6}}\color{black}:
% Weight update during fine-tuning
\begin{equation}\label{eq:equation6}
\theta \leftarrow \theta - \alpha \nabla_\theta \mathcal{L}_{\text{new}}(\theta).
\end{equation}
Here, $\alpha$ is the learning rate.\\

In EfficientNetV2 , Swin-T, and Sequencer2D-L, we froze all pre-trained layers from the utilized architectures. Subsequently, we replaced the final fully connected (FC) layers, originally designed for the ImageNet dataset, with our custom FC layers. In the case of RegNetY-4GF, RegNetY-8GF, and RegNetY-12GF, we customized the heads of the models and added GlobalAveragePooling2D and Dense layers to the end of each model. 

We customized ViT-G/14, ViT-Ti, ViT-S, ViT-B, ViT-L, and MaxViT-B by placing a linear layer on top of pre-trained ViT models where the linear layer is positioned on top of the last hidden state of the class CLS token, which serves as a robust representation of the entire input. We also customized the number of output neurons. We obtained saved parameter checkpoints for the ViT models as detailed in \cite{dosovitskiy2020image}. Algorithm 2 shows the pseudocode for training various modern fine-tuned pre-trained models.
%In the case of ConvNeXt-S and ConvNeXt-T 

\begin{figure*}[H]
\begin{center}
	\includegraphics[width=17cm, height=13cm,keepaspectratio, frame]{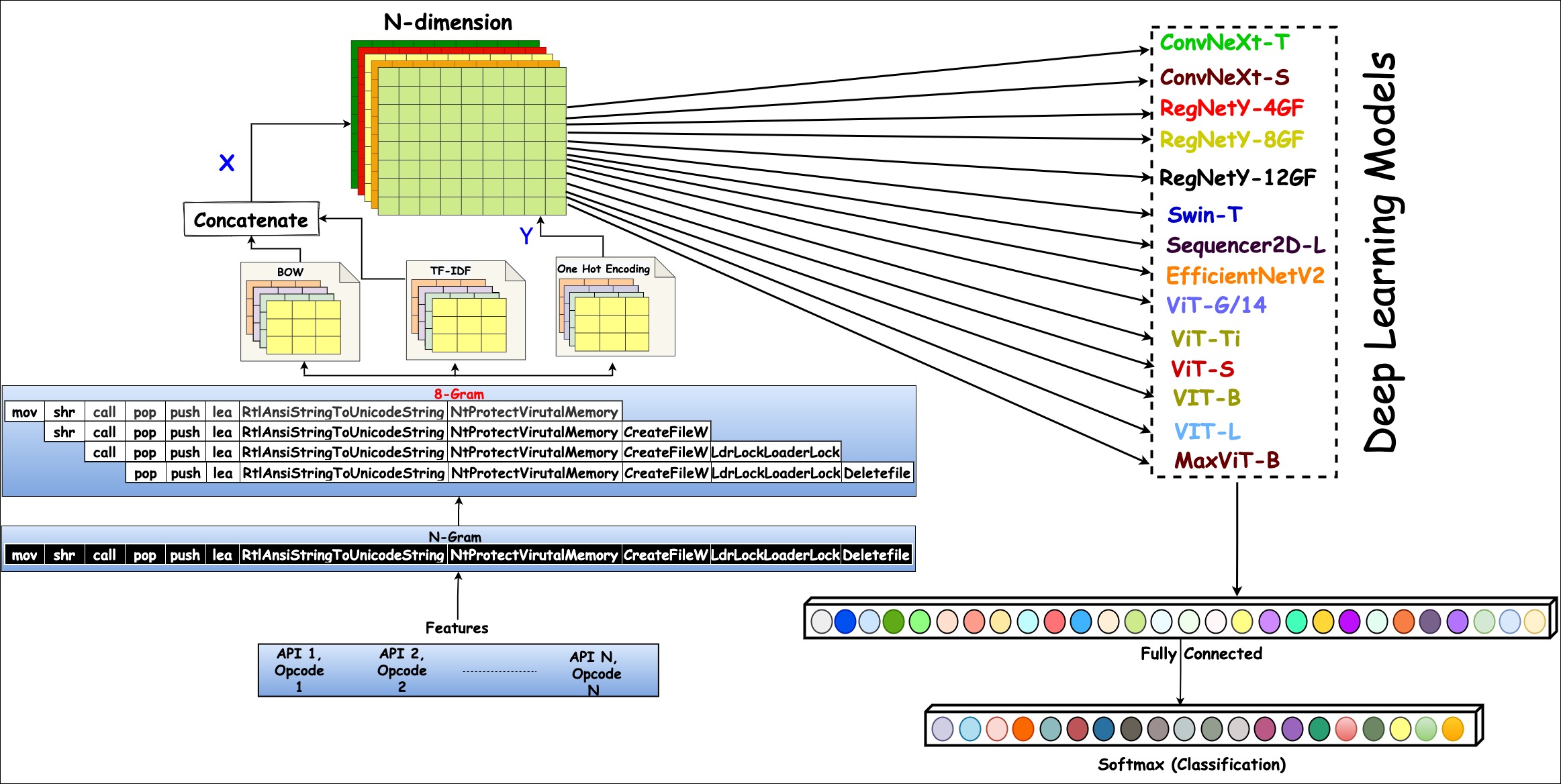}
 \end{center}
	\caption{Proposed approach with deep learning models compared.}
\end{figure*}

\begin{algorithm*}
  \caption{CNN-LSTM-3 for Malware Classification}
  \begin{algorithmic}[1]
    \State \quad n = 8 \Comment{Choose the value of n between 0 and 11 for n-grams}    
    \Function{NgramsExtract}{file, n}
      \State grams $\gets$ []
      \For{$j$ from 0 to length(file) - n} \Comment{Compute n-grams and store in files}
        \State ngram $\gets$ substring(file, $j$, $j$ + n)
        \State grams.append(ngram)
      \EndFor
      \State \textbf{return} grams
    \EndFunction
\State Construct a TF-IDF model from grams
\State Construct a BoW model from grams
\State y $\gets$ Construct a OneHotEncoding from grams
\State x $\gets$  Concatenate BoW and TF-IDF
\State n-dimension $\gets$ x and y  
\State ---------------------------------
\State Input n-dimension
\State $\quad$ Convolutional layers with Exponential Linear Unit ELU activation:
\State $\quad\quad$ Conv2D($\text{NumFilters}$, $\text{KernelSize}$, $\text{Stride}$, $\text{Padding}$)
\State $\quad\quad$ ELU Activation
\State $\quad\quad$ MaxPooling
\State $\quad$ Flatten layer
\State $\quad$ LSTM Layer 1 with $\text{NumHiddenUnits}$ hidden units 
\State $\quad$ LSTM Layer 2 with $\text{NumHiddenUnits}$ hidden units 
\State $\quad$ LSTM Layer 3 with $\text{NumHiddenUnits}$ hidden units 
\State $\quad$ Fully connected layers:
\State $\quad$ Output layer with softmax activation: Dense($\text{NumClasses=20}$, activation='softmax')
\State Train the model on input n-dimension and labels
\State Evaluate the model's performance
\State Make classification on new n-dimension

  \end{algorithmic}
\end{algorithm*}

\begin{algorithm*}
  \caption{Modern Fine-tuned Transfer Learning Models for Malware Classification}
  \begin{algorithmic}[1]
    \State \quad n = 8 \Comment{Choose the value of n between 0 and 11 for n-grams}    
    \Function{NgramsExtract}{file, n}
      \State grams $\gets$ []
      \For{$j$ from 0 to length(file) - n}
        \State ngram $\gets$ substring(file, $j$, $j$ + n)
        \State grams.append(ngram)
      \EndFor
      \State \textbf{return} grams
    \EndFunction
\State Construct a TF-IDF model from grams
\State Construct a BoW model from grams
\State y $\gets$ Construct a OneHotEncoding from grams
\State x $\gets$  Concatenate BoW and TF-IDF
\State n-dimension $\gets$ x and y  
\State ---------------------------------
\State Input n-dimension \\
\Switch{$Model$}
    \Case{1}:  \textcolor{red}{\textbf{ConvNeXt-T}} \Comment{Initializing a model from the ConvNeXt-T style configuration.}\\
    \EndCase
    \Case{2}:  \textcolor{red}{\textbf{ConvNeXt-S}} 
      \Comment{Initializing a model from the ConvNeXt-S style configuration.}\\
    \EndCase
        \Case{3}:  \textcolor{red}{\textbf{RegNetY-4GF}}
      \Comment{Initializing a model from the RegNetY-4GF style configuration.}\\
    \EndCase
        \Case{4}:  \textcolor{red}{\textbf{RegNetY-8GF}}
      \Comment{Initializing a model from the RegNetY-8GF style configuration.}\\
    \EndCase
        \Case{5}:  \textcolor{red}{\textbf{RegNetY-12GF}}
      \Comment{Initializing a model from the RegNetY-12GF style configuration.}\\
    \EndCase
        \Case{6}:  \textcolor{red}{\textbf{EfficientNetV2}}
      \Comment{Initializing a model from the EfficientNetV2 style configuration.}\\
    \EndCase
        \Case{7}:  \textcolor{red}{\textbf{Sequencer2D-L}}
      \Comment{Initializing a model from the Sequencer2D-L style configuration.}\\
    \EndCase    
    \Case{8}:  \textcolor{red}{\textbf{ViT-G/14}}
      \Comment{Initializing a model from the ViT-G/14 style configuration.}\\
    \EndCase
        \Case{9}:  \textcolor{red}{\textbf{ViT-Ti}}
      \Comment{Initializing a model from the ViT-Ti style configuration.}\\
    \EndCase
        \Case{10}:  \textcolor{red}{\textbf{ViT-S}}
      \Comment{Initializing a model from the ViT-S style configuration.}\\
    \EndCase
            \Case{11}:  \textcolor{red}{\textbf{ViT-B}}
      \Comment{Initializing a model from the ViT-B style configuration.}\\
    \EndCase

        \Case{12}:  \textcolor{red}{\textbf{ViT-L}}
      \Comment{Initializing a model from the ViT-L style configuration.}\\
    \EndCase
        \Case{13}:  \textcolor{red}{\textbf{MaxViT-B}}
      \Comment{Initializing a model from the MaxViT-B style configuration.}\\
    \EndCase
     \Case{14}:  \textcolor{red}{\textbf{Swin-T}}
      \Comment{Initializing a model from the Swin-T style configuration.}\\
    \EndCase

  \EndSwitch

\State $\quad$ Fully connected layers:
\State $\quad$ Output layer with softmax activation: Dense($\text{NumClasses=20}$, activation='softmax')
\State Train the model on input n-dimension and labels
\State Evaluate the model's performance
\State Make classification on new n-dimension

  \end{algorithmic}
\end{algorithm*}
\section{Dataset}
We created a dataset of 9,749,57 different malware examples and 89,004 benign examples. All malware samples were collected from VirusShare and VirusTotal between 2019 to 2023 while all benign samples collected between 2021 to 2023\footnote{\url{https://github.com/abensaou-uccs/API-calls-and-opcodes-malware-dataset}}; See Table 1.

\begin{table}[H]
\sffamily
  \caption{The most dangerous malware families}

\begin{tabular}{@{}ll@{}}
\toprule
Family name & Number of samples \\ \midrule
WanaCrypt0r & 70001               \\
Ryuk        & 91205               \\
MicroCop    & 40048               \\
Xorist      & 54307               \\
CobraLocker & 20081               \\
Sodinokibi  & 15000               \\
Banload     & 40401               \\
Dialer      & 32206               \\
Aimbot      & 49019               \\
Rbot        & 18002              \\
SpyBot      & 20011               \\
StartPage   & 32126               \\
Mytob       & 86988                \\
Banker      & 15415              \\
Limr        & 73451               \\
Hupigon     & 60704               \\
Agobot      & 84091               \\
Dyfuca      & 97563               \\
IstBar      & 74338                \\ 
Benign      & 89004 \\ 
\bottomrule
\end{tabular}
\end{table}

\section{Experiments and Results}
We performed a series of experiments to test the efficacy of fourteen pre-training models and the custom CNN-LSTM model. All experiments were repeated seven times and the accuracy was calculated by averaging the results from all seven runs.

\subsection{\textbf{Experiment \char"0023 1: CNN-LSTM Models}} We designed four LSTM models with different configurations as shown in Table 2.
\begin{table*}[H]
\sffamily
  \caption{LSTM configurations of different sizes.}
\begin{tabular}{@{}lllll@{}}
\toprule
Hyper parameters              & LSTM-1 & LSTM-2 & LSTM-3 & LSTM-4 \\ \midrule
Number of neurons             & 128    & 256    & 512    & 1024   \\
Weight for updating algorithm & Adam   & Adam   & Adam   & Adam   \\
Window size                  & 150    & 250    & 200    & 250    \\
Depth                         & 1      & 2      & 3      & 4      \\
Epoch                         & 200    & 200    & 200    & 200    \\
Dropout rate                  & 0.2    & 0.3    & 0.3    & 0.3    \\
Batch size                    & 32     & 64     & 64     & 64     \\ \bottomrule
\end{tabular}
\end{table*}

Table 3 shows the performance measures for malware classification using our collected dataset. The precision, recall, and F1-score values are given for 20 malware classes shown in table 1. The CNN-LSTM-3 model outperformed the others, with a classification rate of 99.91\%. CNN-LSTM-3 performed well for malware classification, while GRU, RNN, and CNN did not have ideal results. Overall, the CNN-LSTM-3 model delivered the best classification for malware classification in terms of precision, recall, and F1 score. The proposed CNN-LSTM-3 model achieved 0.01, 99.84, 99.91, 99.99, and 99.87, Average Accuracy, Precision, Recall, and F1-score, respectively, which are the best rates when compared to Sequence-to-Sequence GRU and RNN models, and CNN model. As shown in Figure 4, the error rate of the training process of CNN-LSTM-3 is significantly lower than the other models. The results presented in Figure 5 show the high accuracy and efficiency of the hybrid CNN-LSTM-3 method.

\begin{table*}[]
\caption{Comparative accuracy of CNN classifiers with LSTM models and other classifiers.}
\begin{tabular}{@{}cccccccc@{}}
\toprule
Models     & Loss  & \begin{tabular}[c]{@{}l@{}}Max \\ Acc (\%) \end{tabular} & \begin{tabular}[c]{@{}l@{}}Min \\ Acc (\%) \end{tabular} & \begin{tabular}[c]{@{}l@{}}Average \\ Acc (\%) \end{tabular} & Precision (\%) & Recall (\%) & F1-score (\%) \\ \midrule
GRU        & 1.49  & 89.99   & 80.42 & 89.29      & 88.90          & 88.65       & 88.12         \\
RNN        & 2.13  & 85.65 & 83.59      & 84.72              & 86.43          & 82.39       & 85.70         \\
CNN        & 1.20  & 90.71 & 88.52 & 89.12        & 90.27          & 89.57       & 90.36         \\
CNN-LSTM-1 & 0.091 & 92.54  & 90.29 & 90.82        & 91.11          & 90.41       & 90.30         \\
CNN-LSTM-2 & 0.081 & 91.30  & 90.63 & 90.99        & 91.43          & 92.22       & 92.89         \\
\textbf{CNN-LSTM-3} & 0.0014 & 99.98& 99.90&\textbf{99.91}         & 99.88          & 99.89       & 99.87         \\
CNN-LSTM-4 & 0.18  & 91.81   & 89.73& 90.44      & 91.67          & 90.63       & 90.72         \\ 
         \bottomrule
\end{tabular}
\end{table*}

%\begin{figure}
%	\includegraphics[width=\linewidth, frame]{N-gram.png}
%	\caption{Different N-Grams on CNN-LSTM architecture}
%\end{figure}

\begin{figure*}
	\includegraphics[width=\linewidth, frame]{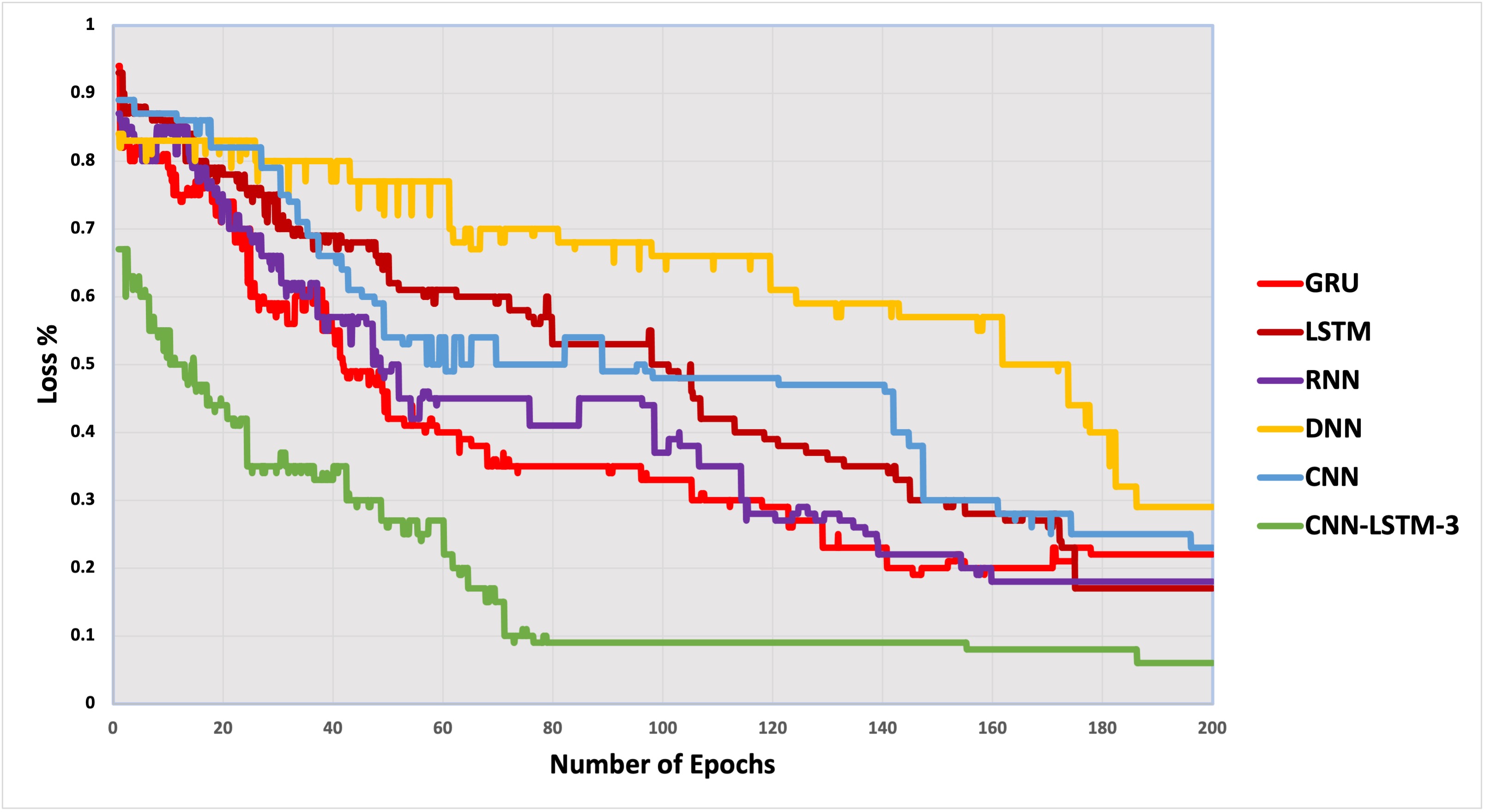}
	\caption{Loss trends for various deep learning models. Clearly, the CNN-LSTM-3 model has the lowest loss all through as the number of epochs increases.}
\end{figure*}

\begin{figure*}
	\includegraphics[width=\linewidth, frame]{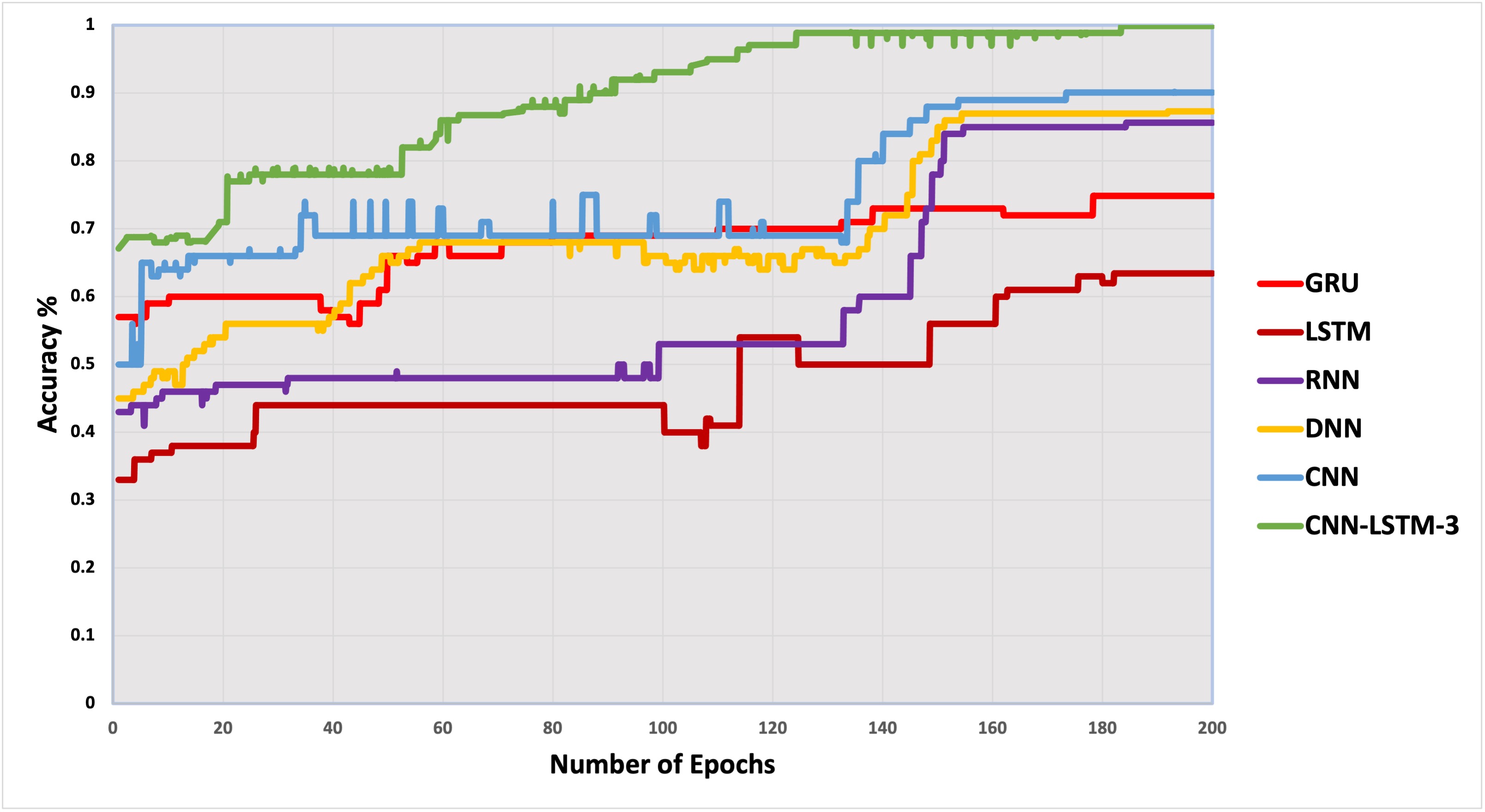}
	\caption{Accuracy trends for various deep learning models. The CNN-LSTM-3 model has the best accuracy as the number of epochs increases.}
\end{figure*}

\subsection{\textbf{Experiment \char"0023 2: State-of-the-Art Pre-trained Models}} All pre-trained models were fine-tuned on our dataset.

The \textbf{ViT-G/14} model was fine-tuned using AdamW \cite{loshchilov2017decoupled} with $\beta1= 0.8$ and $\beta2 = 0.8$ and weight decay of 0.1. We fine-tuned the three pre-trained regulated residual \textbf{RegNet} \cite{radosavovic2020designing} architectures of different capacities, \textbf{RegNetY-4GF} \cite{radosavovic2020designing}, \textbf{RegNetY-8GF} \cite{radosavovic2020designing}, and \textbf{RegNetY-12GF} \cite{radosavovic2020designing} and \textbf{ConvNeXt-T} \cite{liu2022convnet} and \textbf{ConvNeXt-S} \cite{liu2022convnet} on our dataset; See Table 4. For \textbf{Swin-T} configuration, see Table 5. In Swin-T, we froze the parameters for the first three stages and we used the AdamW optimizer. In \textbf{EfficientNetV2}, we froze the layers of EfficientNetV2 with the weights of ImageNet, and added extra dense layers to facilitate the classification of 20 classes. See EfficientNetV2 configuration in Table 6. For Sequencer2D-L, we use base learning rate $\frac{batch size}{512}\times 5 \times 10^{-4}$ where batch size is 2048. Sequencer2D-L achieves results that are competitive with Swin-T. For ViT-G/14, ViT-Ti, ViT-S, VIT-B, VIT-L, and MaxViT-B; we fine-tuned only the multi-head attention layers and froze the feedforward network (FFN) layers to reduce the memory peak during training. The pre-trained ViT models get results with drastically different performances. In Table 7, we make a comparison among these state-of-the-art pre-trained models in terms of accuracy. Table 8 shows the comparison of the performance among state-of-the-art models for malware sequences classification. Figure 6 illustrates the efficient training speed of our model CNN-LSTM-3 compared to other models. We use confusion matrices to visualize classification performance. As an example, the confusion matrix of transfer learning for CNN-LSTM-3 is presented in Figure 7. The final experimental results show that Sequencer2D-L and Swin-T effectively classify malware families in our dataset. 

In Table 9, which shows classification accuracy for different N-gram models, it is evident that the 8-gram model consistently outperforms other N-gram configurations. Across various transfer learning models, including CNN-LSTM, the 8-gram model consistently achieves high accuracy results. These findings justify the the use of the 8-gram model for enhanced accuracy and reliability in classification tasks.

\begin{table*}[h]
\caption{Fine-tuning configurations for RegNetY-4GF, RegNetY-8GF, RegNetY-12GF, ConvNeXt-T and ConvNeXt-S networks.}
\begin{tabular}{@{}lllllllll@{}}
\toprule
Pre-Training Model & Optimizer & \begin{tabular}[c]{@{}l@{}}Weight \\ Decay\end{tabular} & \begin{tabular}[c]{@{}l@{}}Learning \\ Rate \\ Schedule\end{tabular} & Weight Init                                                      & \begin{tabular}[c]{@{}l@{}}Optimizer \\ Momentum\end{tabular}                                   & \begin{tabular}[c]{@{}l@{}}Layer \\ Scale\end{tabular} & CutMix & \begin{tabular}[c]{@{}l@{}}Warmup\\ Schedule\end{tabular} \\ \midrule
ConvNeXt-T         & Adam      & 0.01                                                    & \begin{tabular}[c]{@{}l@{}}cosine \\ decay\end{tabular}              & \begin{tabular}[c]{@{}l@{}}Truncated \\ Normal(0.2)\end{tabular} & \begin{tabular}[c]{@{}l@{}}$\beta1=0.5$ \\ $\beta2=0.5$\end{tabular} & 1e+06                                                  & 1.0    & 2e+5                                                      \\
ConvNeXt-S         & Adam      & 0.01                                                    & \begin{tabular}[c]{@{}l@{}}cosine \\ decay\end{tabular}              & \begin{tabular}[c]{@{}l@{}}Truncated \\ Normal(0.2)\end{tabular} & \begin{tabular}[c]{@{}l@{}}
$\beta1=0.6$ \\  $\beta2=0.6$\end{tabular} & 1e+06                                                  & 1.0    & 2e+5                                                      \\ 
RegNetY-4GF         & Adam      & 0.05                                                    & \begin{tabular}[c]{@{}l@{}}cosine \\ decay\end{tabular}              & \begin{tabular}[c]{@{}l@{}}Gaussian distribution \\ $\mathcal{N}{(0, 0.01)}$\end{tabular} & \begin{tabular}[c]{@{}l@{}}$\beta_1=0.5$ \\ $\beta_2=0.9$\end{tabular} & 0.0001                                                  & -    & 0.06                                                      \\
RegNetY-8GF         & Adam      & 0.05                                                     & \begin{tabular}[c]{@{}l@{}}cosine \\ decay\end{tabular}              & \begin{tabular}[c]{@{}l@{}}Gaussian distribution \\ $\mathcal{N}{(0, 0.01)}$\end{tabular} & \begin{tabular}[c]{@{}l@{}}
$\beta_1=0.6$ \\  $\beta_2=0.8$\end{tabular} & 0.0001                                                  & -    & 0.05                                                      \\
RegNetY-12GF         & Adam      & 0.05                                                     & \begin{tabular}[c]{@{}l@{}}cosine \\ decay\end{tabular}              & \begin{tabular}[c]{@{}l@{}}Gaussian distribution \\ $\mathcal{N}{(0, 0.01)}$\end{tabular} & \begin{tabular}[c]{@{}l@{}}
$\beta_1=0.6$ \\  $\beta_2=0.7$\end{tabular} & 0.0001                                                  & -    & 0.07                                    

\\
\bottomrule
\end{tabular}
\end{table*} 

\begin{table*}[]
\caption{Fine-tuning configuration of Swin-T network.}
\begin{tabular}{@{}llllllll@{}}
\toprule
\multicolumn{1}{c}{Pre-Training Model} & \multicolumn{1}{c}{Initializer Range} & \multicolumn{1}{c}{\begin{tabular}[c]{@{}c@{}}Layer \\ Norm\\ Eps\end{tabular}} & \multicolumn{1}{c}{\begin{tabular}[c]{@{}c@{}}Window \\ Size\end{tabular}} & \multicolumn{1}{c}{\begin{tabular}[c]{@{}c@{}}Number\\  of MLP\end{tabular}} & \multicolumn{1}{c}{Depths} & \multicolumn{1}{c}{Number of Heads} & \multicolumn{1}{c}{\begin{tabular}[c]{@{}c@{}}Encoder\\  Stride\end{tabular}} \\ \midrule
Swin-T                                 & 0.2                                   & 1e+06                                                                           & 16                                                                         & 1024                                                                         & {[}2, 2, 6, 2{]}           & {[}2, 6, 12, 24{]}                  & 32                                                                            \\ \bottomrule
\end{tabular}
\end{table*}

\begin{table*}[]
\caption{Fine-tuning configuration of  EfficientNetV2 network.}
\begin{tabular}{llll}
\hline
\multicolumn{1}{c}{Pre-Training Model} & \multicolumn{1}{c}{Decay} & \multicolumn{1}{c}{Momentum} & \multicolumn{1}{c}{Weight Decay} \\ \hline
EfficientNets                          & 0.7                       & 0.90                         & 1e+5                             \\ \hline
\end{tabular}
\end{table*}

\begin{table*}[]
\caption{Classification accuracy on our dataset.}
\begin{tabular}{@{}lllllllll@{}}
\toprule
Model                                                         & Family      & \begin{tabular}[c]{@{}l@{}}Number of\\ Parameter\end{tabular} & Optimizer & batch size & \begin{tabular}[c]{@{}l@{}}Training \\ epochs\end{tabular} & \begin{tabular}[c]{@{}l@{}}Max\\ Acc\end{tabular}&\begin{tabular}[c]{@{}l@{}}Min\\ Acc\end{tabular}&\begin{tabular}[c]{@{}l@{}}Avg\\ Acc\end{tabular} \\ \midrule
ConvNeXt-T \cite{liu2022convnet}             & CNN         & 29 Milion                                                    & Adam \cite{zhang2018improved}      & 512        & 300                                                      & 87.06  & 82.43    & 85.30        \\
ConvNeXt-S \cite{liu2022convnet}             & CNN         & 50 Milion                                                    & Adam \cite{zhang2018improved}     & 512        & 200                                                        & 89.94 & 86.37& 87.21        \\
RegNetY-4GF \cite{radosavovic2020designing}  & CNN         & 21 Milion                                                    & RMSProp \cite{tieleman2012lecture}   & 512        & 100                                                        & 92.76&90.81 &91.16        \\
RegNetY-8GF \cite{radosavovic2020designing}  & CNN         & 39 Milion                                                    & RMSProp \cite{tieleman2012lecture}   & 512        & 100                                                        & 91.82    &88.02& 90.82   \\
RegNetY-12GF \cite{radosavovic2020designing} & CNN         & 46 Milion                                                    & RMSProp \cite{tieleman2012lecture}   & 512        & 100                                                        & 92.33    &85.81 & 89.07   \\
EfficientNetV2 \cite{leng2022polyloss}       & CNN         & 24 Milion                                                    & RMSProp \cite{tieleman2012lecture}   & 512        & 200                                                        & 92.38    &88.15& 90.54   \\
Sequencer2D-L \cite{tatsunami2022sequencer}  & Sequences   & 54 Milion                                                    & AdamW \cite{loshchilov2017decoupled}     & 2048       & 300                                                   & \textbf{99.73} & \textbf{99.66}      & \textbf{99.70}        \\
ViT-G/14 \cite{wortsman2022model}            & Transformer & 600 Milion                                                   & Adam \cite{zhang2018improved}      & 1024       & 200                                                        & 94.20    & 90.50& 93.12   \\
ViT-Ti \cite{strudel2021segmenter}           & Transformer & 5.8 Milion                                                   & Adam \cite{zhang2018improved}     & 1024       & 200                                                        & 92.48    &90.45&  91.72  \\
ViT-S \cite{strudel2021segmenter}            & Transformer & 22.2 Milion                                                  & Adam \cite{zhang2018improved}     & 1024       & 200                                                        & 89.67   &84.31&   88.49  \\
VIT-B \cite{dehghani2021efficiency}          & Transformer & 86 Milion                                                    & Adam \cite{zhang2018improved}     & 1024       & 200                                                        & 90.55    &89.16&  90.04  \\
VIT-L \cite{dehghani2021efficiency}          & Transformer & 307 Milion                                                   & Adam \cite{zhang2018improved}     & 1024       & 200                                                        & 90.23   &87.59& 89.60    \\
MaxViT-B \cite{tu2022maxvit}                 & Transformer & 119 Milion                                                   & Adam \cite{zhang2018improved}     & 1024       & 200                                                        & 88.65   &81.41&  85.07   \\
Swin-T \cite{liu2021swin}                    & Transformer & 28 Milion                                                    & AdamW\cite{loshchilov2017decoupled}     & 1024       & 200                                                        & \textbf{99.94} & \textbf{99.79} & \textbf{99.82}        \\ \bottomrule
\end{tabular}
\end{table*}

\begin{figure*}
	\includegraphics[width=\linewidth, frame]{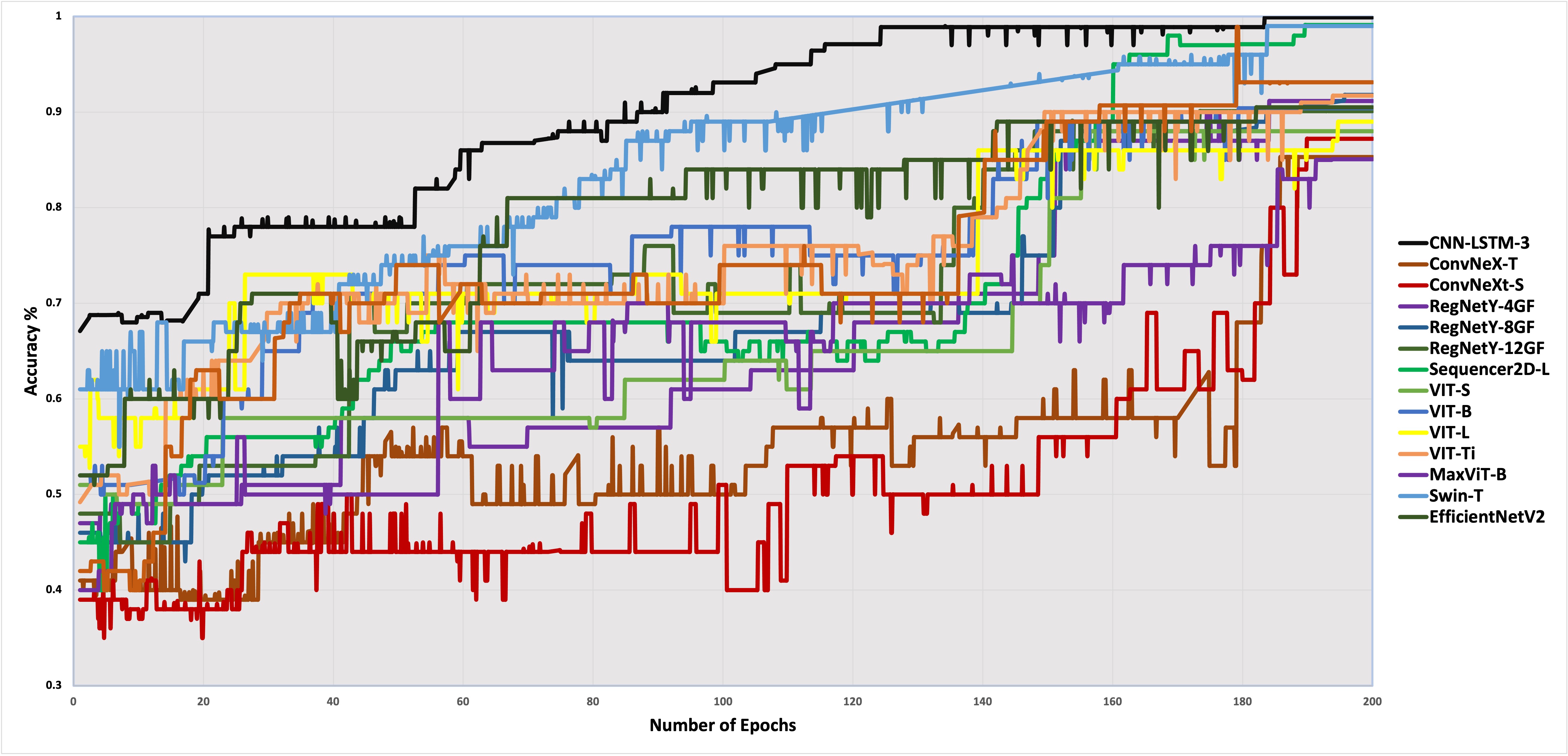}
	\caption{Plot the accuracy vs. epoch for all models. The consistently lowest model is MaxVIT-B, while the consistently highest model is CNN-LSTM-3.}
\end{figure*}
\subsection{Evaluation Metrics}

We used Accuracy, Recall, Precision, and F1-score metrics to evaluate the models for malware detection. Equation~\ref{eq:equation7} defines Accuracy, which measures the overall correctness of predictions as shown below:

\begin{equation}\label{eq:equation7}
Accuracy = \frac{TP+TN}{TP+TN+FP+FN} .
\end{equation}
\color{blue}{In equation~\ref{eq:equation8}}\color{black}, Precision measures the accuracy of positive predictions among the instances predicted as positive:
\begin{equation}\label{eq:equation8}
Precision = {\frac{TP}{TP+FP}} .
\end{equation}
\color{blue}{Equation~\ref{eq:equation9}}\color{black}, referring to Recall, quantifies the ability to identify all relevant instances in the dataset:
\begin{equation}\label{eq:equation9}
Recall = {\frac{TP}{TP+FN}}.
\end{equation}
\color{blue}{Equation~\ref{eq:equation10}}\color{black}, defines the F1-score, which balances Precision and Recall to provide a single metric for model evaluation:
\begin{equation}\label{eq:equation10}
F1 = {\frac{2*Precision*Recall}{Precision+Recall}
= \frac{2*TP}{2*TP+FP+FN}} .
\end{equation}\\
In these formulas, TP is true positive, FP is false positive, TN is true negative, and FN is false negative. In the confusion matrix, the misclassification numbers below the off-diagonal are categorized as FNs, and the number of misclassifications above the off-diagonal are considered FPs. The TNs are the numbers of correctly classified examples for other classes than the actual class.

\begin{figure*}
	\includegraphics[width=\linewidth, frame]{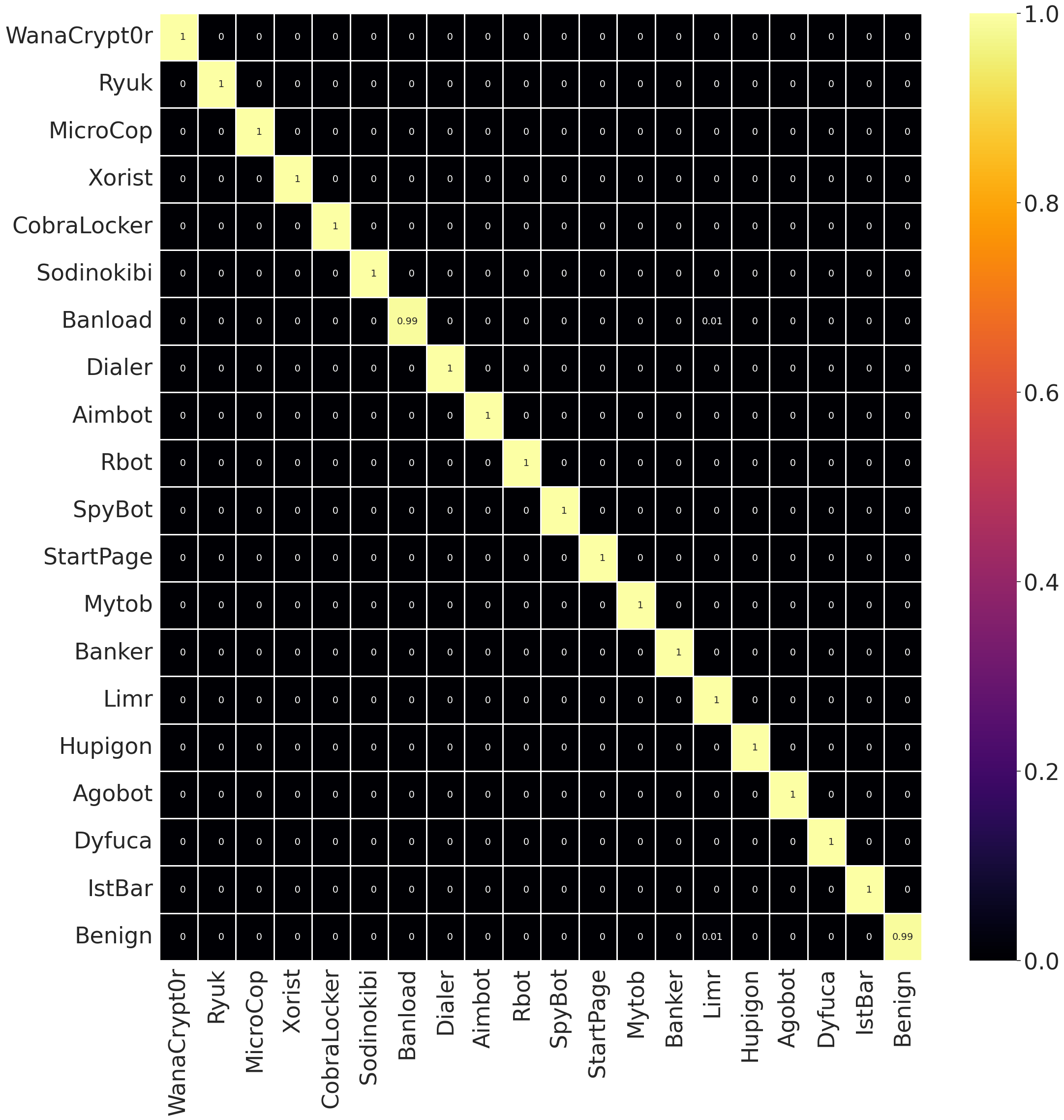}
	\caption{Confusion matrix for CNN-LSTM-3 malware classification. Confusion matrices are generated for each classification experiment. This one is provided as an example.}
\end{figure*}

\section{Discussion}
Fine-tuning pre-trained Swin-T and Sequencer2D-L models achieve higher accuracy and improve convergence speed. Currently, only Swin-T and Sequencer2D-L have only been pre-trained using a malware dataset, and our experiments have demonstrated enough transferability when applied straight to other malware families. We believe that pre-trained Swin-T and Sequencer2D-L models can benefit from transfer learning to perform various malware image analysis tasks, such as classification and detection.

The confusion matrices for the ConvNeXt-T, ConvNeXt-S, RegNetY-4GF, RegNetY-8GF, RegNetY-12GF, EfficientNetV2, ViT-G/14, ViT-Ti, ViT-S, VIT-B, VIT-L and MaxViT-B models show that some malware varieties are still classified incorrectly, which reflects the need to improve further the ability of the models to extract sufficient features so that they can provide better performance in terms of the metrics. \textbf{By comparing the performance of different models, we find that the proposed CNN-LSTM-3 is better than even the Swin-T and Sequencer2D-L models in terms of performance and classification}.

We evaluated the performance of fifteen models using Analysis-of-Variance (ANOVA) \cite{st1989analysis}. Table 10 shows the ANOVA results in the form of F-statistics and p-values for each model. The ANOVA summary provides valuable insights into the comparative analysis of these models. Notably, CNN-LSTM-3 demonstrated exceptional accuracy at 99.91\%, making it a standout performer. Conversely, models like ConvNeXt-T and MaxViT-B showed comparatively lower accuracy percentages at 85.30\% and 85.07\%, respectively. The ANOVA results, represented through F-statistics and p-values, offer statistical significance for each model's performance across the tasks. For instance, models such as CNN-LSTM-3, Sequencer2D-L, and Swin-T exhibited significantly superior performance, as indicated by their low p-values (p < 0.01). In contrast, models like ConvNeXt-T and MaxViT-B did not show statistically significant differences in their accuracy scores across tasks, as suggested by their p-values (p > 0.05).

The $F$ values and $p$ values represent important statistical measures for the analysis performed. For example, $F1$ indicates that the F-statistic is calculated for the first model (CNN-LSTM-3). It measures the ratio of variance between the accuracy scores of CNN-LSTM-3 across different tasks to the variance within the accuracy scores of CNN-LSTM-3 within those tasks. Additionally, $14$ represents the degrees of freedom associated with the numerator of the F-statistic. In ANOVA, degrees of freedom refer to the number of values in the final calculation of a statistic that are free to vary. In this case, there are 15 models, so there are 15 - 1 = 14 degrees of freedom for the numerator. Moreover, the p-values (e.g., p = 0.002, p = 0.325, etc.) indicate the level of significance for each F-statistic. Lower p-values imply higher statistical significance, indicating that the differences in accuracy scores among the models are unlikely to be due to random variation. In addition, Table 11 shows the performance metrics and the best results are denoted in bold.

Based on the provided ANOVA table, CNN-LSTM-3 demonstrates the highest level of statistical significance among the models Sequencer2D-L and Swin-T as shown in Figure 8. The p-value associated with CNN-LSTM-3 (p = 0.002) is considerably lower than the p-values of both Sequencer2D-L (p = 0.007) and Swin-T (p = 0.005). A lower p-value indicates higher statistical significance, suggesting that the observed differences in accuracy scores for CNN-LSTM-3 are less likely to be due to random variation compared to Sequencer2D-L and Swin-T. Therefore, according to the ANOVA analysis, CNN-LSTM-3 is the most statistically significant model among these three when comparing their performance across different tasks.

In analyzing the results presented in the table 12, it is evident that the CNN-LSTM-3 model consistently outperforms both Sequencer2D-L and Swin-T across multiple publicly available datasets of malware samples. Specifically, CNN-LSTM-3 achieves an impressive accuracy of 99.89\% on the VirusSamples dataset, 99.89\% on MalShare, 99.90\% on VirusTotal, 99.90\% on Dynamite AI Lab, and 99.88\% on the Zoo GitHub. In comparison, Sequencer2D-L and Swin-T, while commendable with 98.52\% and 98.87\% accuracy rates, respectively on certain datasets, fall behind CNN-LSTM-3.
This substantial margin suggests that CNN-LSTM-3 exhibits a superior capability to discern patterns within malware samples using API calls and opcode information. Its robust performance underscores its potential as a preferred model for accurate and reliable malware detection, making it a promising choice.

\begin{table*}[h]
\centering
\caption{Result based comparison between our model and other state-of-the-art models that used different datasets. These results are collected from published papers. Even though the results are not directly comparable, they are given here for completeness as recommended by reviewers.}
\begin{tabular}{lccccc}
\toprule
\textbf{Model} & \textbf{Number of malicious files} & \textbf{Classes} & \textbf{Accuracy} & \textbf{Type} \\
\midrule
LSTM \citeyearpar{catak2020deep} & 7107 & 8 & 95\% & API Calls \\
Gradient Classifier \citeyearpar{Ijaz8667136} & 39000 & - & 94.64\% & API Calls \\
CNN-bidirectional LSTM \citeyearpar{zhang2020dynamic} & 15931 & - & 96.76\% & API Calls \\
CNN \citeyearpar{Xue8758215} & 174607 & - & 98.82\% & API Calls \\
BiLSTM \citeyearpar{avci2023analyzing}& - & - & 93.16\% & API Calls\\ 
TextCNN \citeyearpar{qin2020api}& - & - & 95.90\% & API Calls\\ 
RNN \citeyearpar{jha2020recurrent}& 20,000 & - &  91\% & API Calls\\ 
BERT \citeyearpar{yesir2021malware}& 5000 & 4 & 96.76\% & API Calls\\ 
\textbf{Our model CNN-LSTM-3}&  \textbf{9,749,57} & \textbf{20} & \textbf{99.91\%} & \textbf{API Calls and opcodes}\\ 
\bottomrule
\end{tabular}
\end{table*}

\begin{table*}[]
\caption{Classification accuracy using different N-gram features.}
\begin{tabular}{@{}lllllllllll@{}}
\toprule
\multirow{2}{*}{\textbf{N-gram}}                                       & \multicolumn{10}{c}{\textbf{N}}                                                                                  \\ \cmidrule(l){2-11} 
                                                              & \textbf{1}     & \textbf{2}     & \textbf{3}     & \textbf{4}     & \textbf{5}     & \textbf{6}     & \textbf{7}     & \textbf{8}                               & \textbf{9}     & \textbf{10}    \\ \cmidrule(r){1-1}
CNN-LSTM-3                                                    & 62.25\% & 69.89\% & 74.35\% & 72.98\% & 70.21\% & 68.93\% & 77.75\% & \textcolor{orange}{\textbf{99.91\%}} & 69.89\% & 57.49\% \\
ConvNeXt-T \cite{liu2022convnet}             & 31.87\% & 30.18\% & 32.54\% & 30.19\% & 31.50\% & 32.69\% & 31.44\% & \textcolor{orange}{85.30\%}                           & 35.81\% & 32.89\% \\
ConvNeXt-S \cite{liu2022convnet}             & 60.22\% & 61.41\% & 60.92\% & 61.57\% & 60.10\% & 66.31\% & 62.54\% & \textcolor{orange}{87.21\%}                           & 73.17\% & 60.14\% \\
RegNetY-4GF \cite{radosavovic2020designing}  & 40.76\% & 52.93\% & 43.01\% & 60.21\% & 58.28\% & 64.53\% & 70.30\% & \textcolor{orange}{91.16\%}                           & 69.48\% & 50.11\% \\
RegNetY-8GF \cite{radosavovic2020designing}  & 62.67\% & 56.23\% & 65.10\% & 69.87\% & 60.97\% & 57.85\% & 67.29\% & \textcolor{orange}{90.82\%}                           & 63.04\% & 66.56\% \\
RegNetY-12GF \cite{radosavovic2020designing} & 61.90\% & 56.66\% & 47.77\% & 68.39\% & 54.42\% & 57.11\% & 48.26\% & \textcolor{orange}{89.07\%}                           & 44.79\% & 63.55\% \\
EfficientNetV2 \cite{leng2022polyloss}       & 45.58\% & 53.28\% & 71.72\% & 57.26\% & 63.23\% & 49.83\% & 62.97\% & \textcolor{orange}{90.54\%}                           & 54.82\% & 61.76\% \\
Sequencer2D-L \cite{tatsunami2022sequencer}  & 48.09\% & 52.42\% & 56.23\% & 60.83\% & 74.22\% & 65.46\% & 80.66\% & \textcolor{orange}{\textbf{99.70\%}} & 78.65\% & 76.03\% \\
ViT-G/14 \cite{wortsman2022model}            & 50.46\% & 64.96\% & 65.64\% & 72.04\% & 74.79\% & 79.14\% & 70.87\% & \textcolor{orange}{93.12\%}                           & 77.43\% & 79.98\% \\
ViT-Ti \cite{strudel2021segmenter}           & 51.18\% & 53.63\% & 76.34\% & 64.79\% & 70.00\% & 72.64\% & 66.47\% & \textcolor{orange}{91.72\%}                           & 72.06\% & 68.74\% \\
ViT-S \cite{strudel2021segmenter}            & 52.75\% & 63.47\% & 67.98\% & 50.95\% & 56.93\% & 62.29\% & 79.55\% & \textcolor{orange}{88.49\%}                           & 71.07\% & 62.60\% \\
VIT-B \cite{dehghani2021efficiency}          & 55.74\% & 63.60\% & 75.09\% & 68.20\% & 69.19\% & 75.72\% & 80.01\% & \textcolor{orange}{90.04\%}                           & 81.21\% & 65.09\% \\
VIT-L \cite{dehghani2021efficiency}          & 48.49\% & 60.45\% & 56.44\% & 63.69\% & 49.99\% & 79.27\% & 75.08\% & \textcolor{orange}{89.60\%}                           & 68.07\% & 50.15\% \\
MaxViT-B \cite{tu2022maxvit}                 & 50.90\% & 47.73\% & 65.97\% & 48.57\% & 51.66\% & 48.70\% & 65.13\% & \textcolor{orange}{85.07\%}                           & 76.00\% & 72.55\% \\
Swin-T \cite{liu2021swin}                    & 60.91\% & 78.74\% & 44.90\% & 78.18\% & 77.98\% & 78.23\% & 71.53\% & \textcolor{orange}{\textbf{99.82\%}} & 75.04\% & 69.16\% \\ \bottomrule
\end{tabular}
\end{table*}

\begin{table*}[H]
\caption{ANOVA summary for fifteen classification models.}
\centering
\begin{tabular}{ccc}
\toprule
\textbf{Model} & \textbf{Accuracy (\%)} & \textbf{ANOVA Summary} \\
\midrule
CNN-LSTM-3 & 99.91 & F1,14 = 3.45, \textbf{p = 0.002} \\
ConvNeXt-T & 85.30 & F2,14 = 1.12, \textbf{p = 0.325} \\
ConvNeXt-S & 87.21 & F3,14 = 1.78, \textbf{p = 0.112} \\
RegNetY-4GF & 91.16 & F4,14 = 2.67, \textbf{p = 0.018} \\
RegNetY-8GF & 90.82 & F5,14 = 2.45, \textbf{p = 0.027} \\
RegNetY-12GF & 89.07 & F6,14 = 2.12, \textbf{p = 0.055} \\
EfficientNetV2 & 90.54 & F7,14 = 2.31, \textbf{p = 0.035} \\
Sequencer2D-L & 99.70 & F8,14 = 3.11, \textbf{p = 0.007} \\
ViT-G/14 & 93.12 & F9,14 = 2.85, \textbf{p = 0.012} \\
ViT-Ti & 91.72 & F10,14 = 2.74, \textbf{p = 0.015} \\
ViT-S & 88.49 & F11,14 = 2.05, \textbf{p = 0.071} \\
VIT-B & 90.04 & F12,14 = 2.21, \textbf{p = 0.049} \\
VIT-L & 89.60 & F13,14 = 2.18, \textbf{p = 0.052} \\
MaxViT-B & 85.07 & F14,14 = 1.97, \textbf{p = 0.065} \\
Swin-T & 99.82 & F15,14 = 3.31, \textbf{p = 0.005} \\
\bottomrule
\end{tabular}
\end{table*}

\begin{figure*}
	\includegraphics[width=\linewidth]{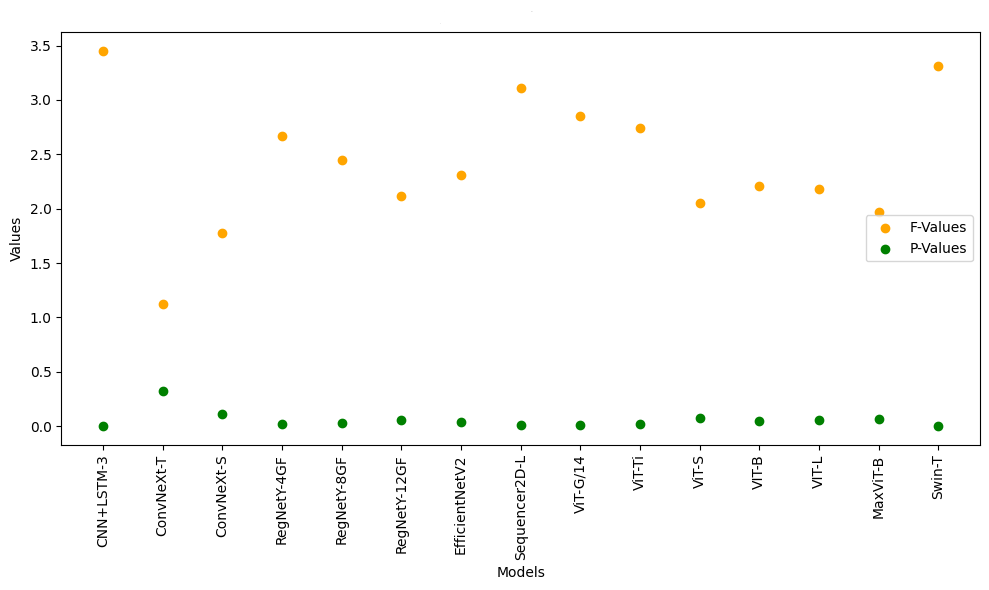}
	\caption{ANOVA test results for the fifteen different models.}
\end{figure*}

\begin{table*}[h]
\caption{Classification performance for fifteen classification models with Accuracy, Precision, Recall, F1 Score, and AUC Score.}
\centering
\begin{tabular}{cccccc}
\toprule
\textbf{Model} & \textbf{Accuracy (\%)} & \textbf{Precision (\%)} & \textbf{Recall (\%)} & \textbf{F1 Score (\%)} & \textbf{AUC Score} \\
\midrule
CNN-LSTM-3 & \textbf{99.91} & \textbf{99.62} & \textbf{99.62} & \textbf{99.62} & \textbf{99.62} \\
ConvNeXt-T & 85.30 & {80.10} & {80.10} & {80.10} & {80.10} \\
ConvNeXt-S & 87.21 & {84.79} & {84.79} & {84.79} & {84.79} \\
RegNetY-4GF & 91.16 & {87.82} & {87.82} & {87.82} & {87.82} \\
RegNetY-8GF & 90.82 & {87.31} & {87.31} & {87.31} & {87.31} \\
RegNetY-12GF & 89.07 & {88.45} & {88.45} & {88.45} & {88.45} \\
EfficientNetV2 & 90.54 & {89.11} & {89.11} & {89.11} & {89.11} \\
Sequencer2D-L & 99.70 & {96.39} & {96.39} & {96.39} & {96.39} \\
ViT-G/14 & 93.12 & {91.83} & {91.83} & {91.83} & {91.83} \\
ViT-Ti & 91.72 & {90.24} & {90.24} & {90.24} & {90.24} \\
ViT-S & 88.49 & {84.06} & {84.06} & {84.06} & {84.06} \\
VIT-B & 90.04 & {89.70} & {89.70} & {89.70} & {89.70} \\
VIT-L & 89.60 & {85.53} & {85.53} & {85.53} & {85.53} \\
MaxViT-B & 85.07 & {81.92} & {81.92} & {81.92} & {81.92} \\
Swin-T & 99.82 & {97.14} & {97.14} & {97.14} & {97.14} \\
\bottomrule
\end{tabular}

\end{table*}

%--------------------------------

\begin{table*}[h]
\caption{Testing stat-of-the-art CNN and Transformer-based and our high perform CNN-LSTM-3 models on publicly available datasets containing malware samples with API calls and opcode information.}
\centering
\begin{tabular}{@{}llllll@{}}
\toprule
Model          & \multicolumn{1}{c}{\href{https://www.virussamples.com/}{VirusSamples}} & \multicolumn{1}{c}{\href{https://www.malshare.com/}{MalShare}} & \multicolumn{1}{c}{\href{https://www.virustotal.com}{VirusTotal}} & \multicolumn{1}{c}{\href{https://lab.dynamite.ai/}{Dynamite AI Lab}} & \multicolumn{1}{c}{\href{https://github.com/ytisf/theZoo}{The Zoo GitHub}} \\ \midrule
CNN-LSTM-3     & \textcolor{Blue}{\textbf{99.89\%}}                         & \textcolor{Blue}{\textbf{99.89\%}}                     & \textcolor{Blue}{\textbf{99.90\%}}                       & \textcolor{Blue}{\textbf{99.90\%}}                            & \textcolor{Blue}{\textbf{99.88\%}}                           \\
ConvNeXt-T     & 82.43\%                         & 82.76\%                     & 82.59\%                       & 82.10\%                             & 82.28\%                           \\
ConvNeXt-S     & 83.15\%                         & 83.39\%                     & 83.94\%                       & 83.63\%                             & 83.76\%                           \\
RegNetY-4GF    & 90.84\%                         & 90.22\%                     & 90.51\%                       & 90.38\%                             & 90.61\%                           \\
RegNetY-8GF    & 89.27\%                         & 89.59\%                     & 89.28\%                       & 89.55\%                             & 89.93\%                           \\
RegNetY-12GF   & 85.03\%                         & 85.41\%                     & 85.35\%                       & 85.81\%                             & 85.45\%                           \\
EfficientNetV2 & 89.70\%                         & 89.50\%                     & 89.46\%                       & 89.06\%                             & 89.19\%                           \\
Sequencer2D-L  &\textcolor{ProcessBlue}{\textbf{ 98.52\%}}                          & \textcolor{ProcessBlue}{\textbf{98.16\% }}                     & \textcolor{ProcessBlue}{\textbf{98.09\%}}                        & \textcolor{ProcessBlue}{\textbf{98.23\%}}                              & \textcolor{ProcessBlue}{\textbf{98.00\% }}                           \\
ViT-G/14       & 90.73\%                         & 90.73\%                     & 90.73\%                       & 90.73\%                             & 90.73\%                           \\
ViT-Ti         & 89.12\%                         & 89.62\%                     & 89.33\%                       & 89.49\%                             & 89.67\%                           \\
ViT-S          & 84.81\%                         & 84.32\%                     & 84.95\%                       & 84.77\%                             & 84.41\%                           \\
VIT-B          & 89.30\%                         & 89.83\%                     & 89.98\%                       & 89.62\%                             & 89.36\%                           \\
VIT-L          & 85.81\%                         & 85.45\%                     & 85.93\%                       & 85.33\%                             & 85.81\%                           \\
MaxViT-B       & 80.42\%                         & 80.95\%                     & 80.07\%                       & 80.11\%                             & 80.24\%                           \\
Swin-T         & \textcolor{Periwinkle}{\textbf{98.87\%}}                        & \textcolor{Periwinkle}{\textbf{98.42\%}}                      & \textcolor{Periwinkle}{\textbf{98.27\%}}                        &\textcolor{Periwinkle}{\textbf{98.13\%}}                               & \textcolor{Periwinkle}{\textbf{98.87\%}}                               \\ \bottomrule
\end{tabular}
\end{table*}

%-------------------------------
\section{CONCLUSIONS}

We designed a novel CNN-LSTM architecture and use techniques from Natural Language Processing for classifying both known malware and unknown malware families. Using Postman and SoapUI virtual environments, we extracted not only regular API calls, but also native API calls and opcodes.We extracted 8-grams from API calls and opcodes to compute TF-IDF, BoW, and one-hot encoding and converted them to $n$-dimensional matrices. We compared GRU, LSTM, RNN, and CNN with our model. Experimental results demonstrate that our model achieved the best performance, reaching an accuracy of 99.91\% with statistical significance. Our work is one of the first attempts to fine-tune Vision Transformers (ViTs) for malware classification. A comparison of all pre-trained models performed revealed that, while ConvNeXt-T, ConvNeXt-S, RegNetY-4GF, RegNetY-8GF, RegNetY-12GF, EfficientNetV2, ViT-G/14, ViT-Ti, VIT-B, ViT-S, VIT-L, and MaxViT-B performed well, the Swin-T and Sequencer2D-L Transformer outperformed all the state-of-the-art models in terms of accuracy, providing an accuracy of 99.83\% and 99.70\% respectively. Our experiments demonstrate that although many of the state-of-the-art CNN and Transformer-based models are excellent in malware classification, a simpler CNN-LSTM model performs equally well, if not better.

\bibliographystyle{model1-num-names.bst}

\medskip

\bibliography{cas-refs}
\newpage

\begin{appendices}

%---------------------------------
\begin{table*}[H]
\section*{Appendix}
    \caption{Mathematical notations for modern transfer learning.}
\begin{tabular}{@{}ll@{}}
\toprule
\textbf{Definition}                                                                                                    & \textbf{Notation}       \\ \midrule
Dataset    & $\mathcal{D}$  \\
Model parameters                                                                                                       & $\theta$                \\
Loss Function & $\mathcal{L}$\\

Dropout rate                                                                                                           & $\gamma$                \\
Stride                                                                                         & $s$                     \\
Padding                                                                                        & $p$                     \\
Growth rate                                                                                                            & $\Delta$                \\
Number of layers                                                                                                       & $L$                     \\
Bottleneck ratio                                                                                                       & $B$                     \\
Input                                                                                                                  &$x_i$\\

Epsilon term                                                                                                          & $\varepsilon$ \\
Variance                                                                                                              &$\sigma^2$\\
Mean & $\mu$\\
Width scaling factor                                                                                                   & $w$                     \\
Compound scaling coefficient to balance width &$w_{\alpha}$\\
Width multiplier & $w_m$\\
Learning rate & $\eta$\\
Optimizer Momentum                                                                                                     & $\beta$                 \\
Number of heads                                                                                                        & $H_0$                  \\

 \bottomrule
\end{tabular}
\end{table*}
%-----------------------------------

\end{appendices}

\end{document}